\documentclass[review, 1p, sort&compress]{elsarticle}

\usepackage[utf8]{inputenc}
\usepackage[T1]{fontenc}
\usepackage{mhchem}

\journal{Chemical Physics}

\begin{document}

\begin{frontmatter}

\title{Solvation Effects Alter the Photochemistry of 2-Thiocytosine}



\author[pwr]{Mikołaj J. Janicki}
\author[ibp,ifpan]{Rafał Szabla\corref{corresponding}}
\ead{szabla@ifpan.edu.pl}
\author[ibp]{Jiří Šponer}
\author[pwr]{Robert W. Góra\corref{corresponding}}
\ead{robert.gora@pwr.edu.pl}
\cortext[corresponding]{Corresponding author}



\address[pwr]{Department of Physical and Quantum Chemistry, Faculty of
Chemistry, Wroclaw University of Science and Technology, Wybrzeże
Wyspiańskiego 27, 50-370, Wrocław, Poland}
\address[ibp]{Institute of Biophysics of the Czech Academy of Sciences,
Královopolská 135, 61265 Brno, Czech Republic}
\address[ifpan]{Institute of Physics, Polish Academy of Sciences, Al.
Lotników 32/46, 02-668 Warsaw, Poland}

\begin{abstract}
Radiationless deactivation channels of 2-thiocytosine in aqueous
environment are revisited by means of quantum-chemical simulations
of excited-state absorption spectra, and investigations of potential
energy surfaces of the chromophore clustered with two water
molecules using the algebraic diagrammatic construction method to
the second-order (ADC(2)), and multireference configuration
interaction with single and double excitations (MR-CISD) methods. We
argue that interactions of explicit water molecules with
thiocarbonyl group might enable water-chromophore electron transfer
(WCET) which leads to formation of intersystem crossing that was not
considered previously. This is the first example of a WCET process
occurring in the triplet manifold of electronic states. This
phenomenon might explain nonradiative decay of the triplet state
population observed in thiopyrimidines in the absence of molecular
oxygen. According to our calculations this WCET process might also
entail a subsequent, virtually barrierless, electron-driven proton
transfer (EDPT) resulting in the formation of hydroxyl radical which
could further participate in photohydration or deamination
reactions.
\end{abstract}


\end{frontmatter}


\section{Introduction}
Photochemistry and photophysics of thiated nucleobases has recently
attracted increasing attention, owing to to their potential use as
photosensitizers in pharmacological applications
\cite{arslancan.m.v22.2017.p998}, nanotechnology
\cite{wojciechowski.alternative.v40.2011.p5669}, and their
intriguing properties in promoting prebiotically credible chemical
reactions and processes \cite{szostak.jsc.v3.2012.p2,
zhang.pnas.v110.2013.p17732, heuberger.jacs.v137.2015.p2769,
weiss.nm.v1.2016.p16116, xu.nc.v9.2017.p303}.  For instance,
$\alpha$-2-thioribocytidine was recently demonstrated to be a
crucial intermediate in the photochemical synthesis of
$\beta$-ribonucleosides under the conditions of the early Earth
\cite{xu.nc.v9.2017.p303}. The prebiotic background of thiated
nucleobases is also supported by presence of naturally occurring
thiobases in mitochondrial tRNA of some bacteria and yeasts
\cite{carbon.s.v161.1968.p1146, ajitkumar.mr.v52.1988.p103,
cherayil.crc.1990}.
However, the main focus on thiated nucleobases stems from their
potential applications in photochemotherapy
\cite{attard.uva.v11.2011.p62,pridgeon.bjc.v104.2011.p1869,
pollum.increase.v17.2015.p27851}, photodynamic therapies in treating
skin disorders \cite{reelfs.p&ps.v11.2011.p148}, superficial tumors
\cite{massey.cb.v11.2001.p1142, zhang.dr.v6.2007.p344} and bladder
cancer \cite{pridgeon.bjc.v104.2011.p1869}. It is worth noting that
thioguanine is already used as a therapeutic agent for lymphoblastic
leukemia \cite{vora.l.v368.2006.p1339} and breast cancer
\cite{johnson.ar.v31.2011.p2705}, while 2-thiocytosine influences
the mitosis of human lymphocytes
\cite{robin_t._b._rye.mass.v62.1984} and shows cytotoxic
\cite{lozzio.ecr.v69.1971.p377}, anticancer
\cite{vetter.ica.v362.2009.p189} and antibacterial activity
\cite{fillat.ejic.v2011.2011.p1487}.

Characterization of photochemical and photophysical properties of
thiated nucleobases is crucial for understanding the mechanisms
governing their phototherapeutic activity and photoinduced chemical
reactions. In the past decade, various spectroscopic and theoretical
studies aimed to scrutinize the photophysics and photochemistry
of thiocytosine, thioguanine, thiouracil, and thiothymine
\cite{harada.jpcb.v111.2007.p5518, reichardt.jpcl.v1.2010.p2239,
reichardt.cc.v46.2010.p5963, reichardt.jpcb.v115.2011.p3263,
martinez-fernandez.cc.v48.2012.p2134, cui.jcp.v138.2013.p044315,
martinez-fernandez.cs.v5.2014.p1336,
taras-goslinska.jpapac.v275.2014.p89,
gobbo.catc.v10401041.2014.p195, pollum.increase.v17.2015.p27851,
mai.jpca.v119.2015.p9524, mai.nc.v7.2016.p13077,
mai.jpcl.v7.2016.p1978}. It is now well established that the most
attractive features of thionucleobases are their structural
similarity to canonical nucleobases combined with considerable
singlet-triplet spin-orbit couplings (SOC) that enables efficient
population of long-lived triplet states and efficient generation of
singlet oxygen \cite{pollum.tcc.v355.2015.p245,
arslancan.m.v22.2017.p998}. Recently joint transient absorption
spectroscopy (TAS) and computational studies of Mai \textit{et al.}
\cite{mai.nc.v7.2016.p13077} revealed that near unity triplet yields
often observed in thionucleobases may also be explained by
relatively high energies and consequently lower accessibility of
S$_{1}$/S$_{0}$ conical intersections when compared to the
lowest-energy intersystem crossings.

Although many aspects of the photophysics of thionucleobases are now
quite well understood, previous computational studies focused
predominantly on isolated chromophores. This may be a serious
limitation obscuring interpretation of the experimental results,
since recently there is a growing amount of data suggesting that
water molecules may actively participate in photochemistry of
hydrated heterocycles by modifying state crossings and, more
importantly, opening new photorelaxation channels which are not
accessible in the gas phase. The examples include microhydrated
adenine \cite{szabla.photorelaxation.v195.2017.p237,
barbatti.jacs.v136.2014.p10246,
chaiwongwattana.jpca.v119.2015.p10637, wu.cpc.v17.2016.p1298},
cytosine and cytidine \cite{szabla_waterchromophore_2017},
eumelanine \cite{nogueira_eumelanin_2017}, and azole chromophores
\cite{szabla.pccp.v16.2014.p17617, szabla.jpcl.v4.2013.p2785}. For
instance, Barbatti demonstrated that water molecules may stabilize
the excited $^1n_{\rm N}\pi^{\ast}$ state in 7H-adenine through the
interaction of the $n_{\rm N}$ orbital of adenine with the p$_{z}$
orbital of the neighbouring water molecule
\cite{barbatti.jacs.v136.2014.p10246}. This permits radiationless
deactivation through the $n\pi^{\ast}$/$S_0$ state crossing, which
is induced by \emph{water-to-chromophore electron transfer} (WCET).
More recently, some of us suggested that the long-lived
dark state observed experimentally in aqueous cytidines might
correspond to the $^1n\pi^{\ast}_{CT}$ excited state also involved in
WCET \cite{szabla_waterchromophore_2017}.
Consequently, understanding how the surrounding water molecules
could affect the photochemistry of thionucleobases (\textit{e.g.}
thiocytosine) is the obvious next step in providing their complete
photochemical characteristics in native environments.

In this work, we show that microhydration of 2-thiocytosine by two
water molecules [2tCyt-$\ce{(H2O)2}$] may enable a similar WCET
process, which could lead to the formation of yet another
intersystem crossing in microsolvated 2tCyt that was not considered
previously. This phenomenon might also explain nonradiative decay of
the triplet state population observed in thiopyrimidines in the
absence of molecular oxygen \cite{pollum.tcc.v355.2015.p245}.

\section{Computational methods}
The ground-state minimum energy geometries and harmonic vibrational
frequencies of microhydrated 2tCyt were computed at the MP2/cc-pVTZ
level \cite{weigend_ri-mp2:_1997, weigend_ri-mp2:_1998,
hattig_distributed_2006, kendall_electron_1992}. The stationary
points on excited state potential energy surfaces were located using
the algebraic diagrammatic construction to the second order method
[ADC(2)] method \cite{schirmer_beyond_1982,
hattig_structure_2005,dreuw_algebraic_2015} and the cc-pVTZ basis
set as implemented in the \textsc{Turbomole 7.1} package
\cite{TURBOMOLE}.

Vertical excitation energies were obtained at the ADC(2)/cc-pVTZ
level, assuming the S$_0$ geometry optimized using the MP2/cc-pVTZ
method. Solvation effects on the vertical excitation energies
exerted by bulk water were estimated using the non-equilibrium
polarizable continuum solvation model (PCM) combined with
ADC(2)/cc-pVTZ method which are implemented in the \textsc{Q-Chem
5.0} package \cite{QCHEM4}. The perturbed state specific (ptSS)
approach was used either in a fully consistent
perturbation-energy-and-density (PTED) or perturbation-density (PTD)
variant \cite{mewes.jpca.v119.2015.p5446, mewes_accuracy_2017}. The
orbital character of the considered excited states was assigned
based on natural transition orbitals \cite{martin_natural_2003}
(NTOs) obtained using \textsc{TheoDore} package \cite{theo}. The
potential energy (PE) profiles for WCET mechanism were calculated
using the ADC(2) and MP2 methods for the excited and ground
electronic states, respectively, using the \textsc{Turbomole 7.1}
package. The PE scans were performed based on linear interpolation
in internal coordinates (LIIC) between the appropriate stationary
geometries.

The minimum-energy crossing points (MECPs) were located using the
sequential penalty constrained optimization method proposed by
Levine et al.  \cite{levine.jpcb.v112.2008.p405} employing energies
and analytical gradients computed at the ADC(2) and MP2 levels, for
the excited and ground state, respectively, and assuming the cc-pVTZ
basis set. The MECPs were optimized using \textsc{CIOpt} package
\cite{levine.jpcb.v112.2008.p405} which was interfaced with the
\textsc{Turbomole 7.1} program. This approach was also employed to
locate T$_{2}$($^{3}\pi_{\rm S}\pi^{\ast}$)/T$_{1}$($^{3}n_{\rm
S}\pi^{\ast}$) MECP, which in terms of non-adiabatic transition
state theory \cite{lorquet_nonadiabatic_1988} represents the saddle
point (transition state) separating the ring-puckered and WCET
minima on the T$_1$ hypersurface. The transition state structure was
verified by the assignment of the orbital character of the
degenerated T$_{2}$ and T$_{1}$ states, and the dominant
contributions were associated with the $^{3}\pi_{\rm S}\pi^{\ast}$
(ring-puckered) and $^{3}n_{\rm S}\pi^{\ast}$ (WCET) characters
respectively.

To validate whether the ADC(2) method is appropriate for the studied
case the T$_1$ ($n_{\rm S}\pi^{\ast}$) minimum-energy and
T$_1$/S$_0$ MECI structures of microhydrated 2tCyt obtained using
ADC(2) method were reoptimized at MR-CISD(2,3)/6-31++G(d,p) level
using the \textsc{Columbus 7.0} package
\cite{lischka_columbus_2012}. The reference configurations were
obtained by distributing two electrons among occupied $n_{\rm S}$
and two virtual $\pi^{\ast}$ orbitals. The Davidson size-extensivity
correction was applied to all MR-CISD energies (MR-CISD+Q). 
The spin-orbit coupling (SOC) was computed for the S$_1$/T$_2$ MECP
ISC at the MS-CASPT2/cc-pVTZ-DK level using MOLCAS 8.0 package
\cite{ aquilante_molcas_2016}. The 2nd order Douglas--Kroll--Hess
Hamiltonian was adopted in these computations to include scalar
relativistic effects. 

The excited-state absorption (ESA) spectra of isolated and
microhydrated 2tCyt were simulated employing the nuclear ensemble
method \cite{crespo-otero_spectrum_2012}, with 500 initial
conditions generated from a Wigner distribution for all vibrational
normal modes calculated for the appropriate minimum-energy excited
state structures using the ADC(2) method and the cc-pVTZ basis set.
The required excitation energies and oscillator strengths for the 11
lowest excited states were obtained using ADC(2) method and the same
basis set. The ESA spectra were simulated using the modified
\textsc{Newton-X 2.0} program \cite{barbatti_newton-x:_2014} which
was interfaced with the \textsc{Turbomole 7.1} package.

\section{Results and Discussion}

\subsection{Vertical excitations in the light of UV-absorption
spectrum of 2-thiocytosine}
The stationary absorption spectrum of 2tCyt in water solution shows
a broad band with maxima at 270 nm (4.59 eV), 240 nm (5.17 eV) and a
shoulder at about 220 nm (5.64 eV) \cite{mai_solvatochromic_2017}.
Based on the results of ab initio calculations, the first peek was
assigned by Mai \emph{et al.} to S$_4$ and S$_2$ states having
similar $^1\pi_{\rm S}\pi^{\ast}$ character (centered around the
thiocarbonyl group but associated with different $\pi^{\ast}$
orbitals), and the remaining features to the S$_6$ and  S$_8$
states, respectively, arising from $^1\pi\pi^{\ast}$ transitions
within the aromatic ring \cite{mai_solvatochromic_2017}. The 2tCyt
water cluster used in these calculations was constructed by
saturating all the possible water-chromophore hydrogen bonds, and
the vertical excitation energies were computed using the multi-state
complete active space perturbation theory method (MS-CASPT2)
including the non-equilibrium polarizable continuum (PCM) implicit
solvation model.

\begin{table}
\caption{Vertical excitation energies of microhydrated 2tCyt, computed using
ADC(2)/ptSS-PCM(PTED)/cc-pVTZ approach, assuming the S$_0$ structure optimized
at the MP2/PCM/cc-pVTZ level. The corresponding MS-CASPT2(14,10)/PCM/cc-pVDZ
values were taken from Ref. \citenum{mai_solvatochromic_2017} and are shown
for the reference.}
\label{tab:vertical}
\begin{center}
\begin{tabular}{@{\extracolsep{\fill}}cccccc}
\hline\noalign{\smallskip}
\multicolumn{2}{c}{State / Transition} & \multicolumn{2}{c}{ADC(2)} & \multicolumn{2}{c}{MS-CASPT2} \\
&& E$_{\rm exc}$/[eV] & f$_{osc}$ & E$_{\rm exc}$/[eV] & f$_{osc}$ \\
\noalign{\smallskip}\hline\noalign{\smallskip}
S$_1$  & $n_{\rm S}\pi^{\ast}  $  & 4.00  &  2.1$\cdot10^{-4}$ & 4.50 &  0.02 \\
S$_2$  & $\pi_{\rm S}\pi^{\ast}$  & 4.30  &  0.10              & 4.23 &  0.01 \\
S$_3$  & $n_{\rm S}\pi^{\ast}  $  & 4.40  &  1.7$\cdot10^{-3}$ & 5.12 &  0.02 \\
S$_4$  & $\pi_{\rm S}\pi^{\ast}$  & 4.52  &  0.30              & 4.60 &  0.40 \\
\noalign{\smallskip}\hline\noalign{\smallskip}
T$_1$  & $\pi_{\rm S}\pi^{\ast}$  & 3.73  &   -                &  -   &  -    \\
T$_2$  & $n_{\rm S}\pi^{\ast}  $  & 3.93  &   -                &  -   &  -    \\
T$_3$  & $\pi_{\rm S}\pi^{\ast}$  & 4.00  &   -                &  -   &  -    \\
\noalign{\smallskip}\hline\noalign{\smallskip}
\end{tabular}
\end{center}
\end{table}

\begin{figure}
\centering\includegraphics[width=1\linewidth]{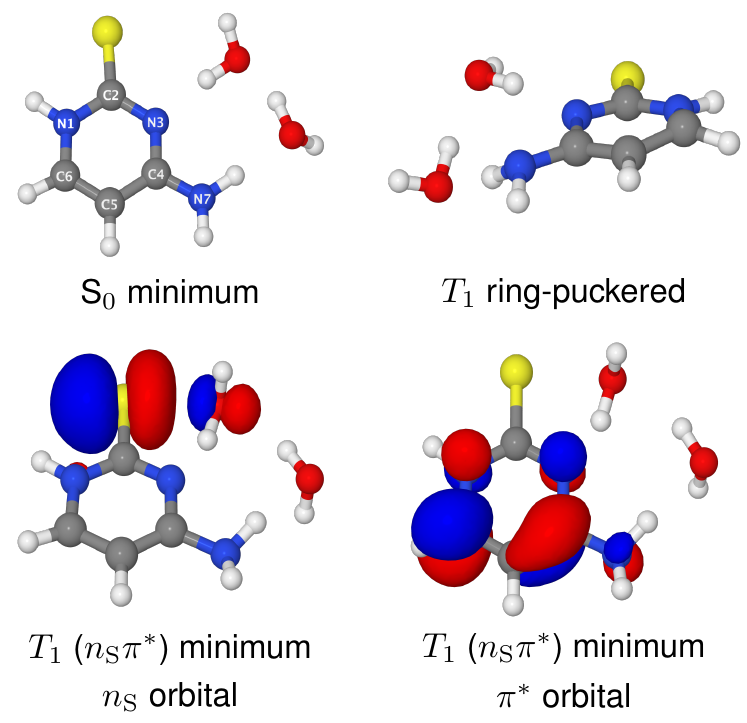}
\caption{The geometries of microhydrated 2tCyt optimized using the
ADC(2)/cc-pVTZ method. The n$_{\rm S}$ and $\pi^{\ast}$ molecular orbitals
indicate WCET character of lowest-lying T$_1$ triplet state.}
\label{fig:geoms}
\end{figure}

The vertical excitations of the 2tCyt-$\ce{(H2O)2}$ cluster
considered in present contribution (cf. Fig. \ref{fig:geoms}),
computed using the ADC(2)/cc-pVTZ method with non-equilibrium
ptSS-PCM implicit solvation model of bulk water in a fully
self-consistent PTED variant, are shown in Tab.~\ref{tab:vertical}.
The corresponding gas-phase data are shown in Tab.~S1 in the
Supplementary Information (SI). Generally, our results are in
agreement with earlier MS-CASPT2 predictions with one notable
exception. In both ADC(2) and MS-CASPT2 calculations the low-lying
$^1\rm{n}_{\rm S}\pi^{\ast}$ and $^1\pi_{\rm S}\pi^{\ast}$ states
are destabilized in polarizable medium but to different extent. That
is why we do not observe the change of ordering of S$_1$ and S$_2$
states in our ADC(2) results that is predicted by the MS-CASPT2/PCM
approach \cite{mai_solvatochromic_2017}. It is worth noting that the
reaction field in MS-CASPT2 calculations was computed at the CASSCF
level, thus lacking the dynamic correlation effects, while the
ADC(2)/ptSS-PCM(PTED) calculations were self-consistent with the
correlated electron density. Our assumption is further confirmed by
the fact that the ordering of the  $^1\rm{n}_{\rm S}\pi^{\ast}$ and
$^1\pi_{\rm S}\pi^{\ast}$ states is exchanged only in the MS-CASPT2
calculations including the PCM solvation model, while inclusion of
five explicit water molecules without polarizable dielectric
environment in the MS-CASPT2 calculation yielded a result which is
in qualitative agreement with the ADC(2)/ptSS-PCM(PTED) calculations
\cite{mai_solvatochromic_2017}.

Since the ordering of states is important for further discussion, we
decided to reoptimize the structure with five explicit water
molecules from Ref. \citenum{mai_solvatochromic_2017} at the
MP2/PCM/cc-pVTZ level. This geometry was further used to compute
vertical excitation energies using the ADC(2)/ptSS-PCM(PTED)/cc-pVTZ
approach. The results indicate that the $^1n_{\rm S}\pi^{\ast}$
transition is still the lowest-lying state (4.40 eV) and the second
one is the $^1\pi_{\rm S}\pi^{\ast}$ state (4.54 eV), albeit the
energy difference between these excitations has decreased. We also
performed calculations with the same cluster surrounded by more than
200 water molecules represented by the effective fragment potentials
(EFP) and implicit solvent model at the
ADC(2)/EFP/ptSS-PCM(PTD)/cc-pVDZ level which also show the same
ordering of the low-lying states (cf. SI for details).

\subsection{Nonradiative deactivation channels of microsolvated
2-thiocytosine}
The experimental transient absorption spectra (TAS) recorded for 2tCyt show
broad and featureless absorption during the early \mbox{100 fs} which was
tentatively assigned to population of singlet $\pi_{\rm S}\pi^{\ast}$ and
$n_{\rm S}\pi^{\ast}$ states \cite{mai.nc.v7.2016.p13077}. After the initial
120 fs, the spectrum exhibits two distinct absorption maxima at about 350 and
550 nm.  The former maximum is decaying after 320 fs while the latter is
continually rising during the first 3.0 ps. These bands were assigned to the
singlet $^1\pi_{\rm S}\pi^{\ast}$ and triplet $^3\pi_{\rm S}\pi^{\ast}$
states, respectively (possibly with some contribution from $^3n_{\rm
S}\pi^{\ast}$ transition), and the latter peak does not decay within 20 ps
which confirms its assignment to the long-lived triplet state
\cite{arslancan.m.v22.2017.p998}.
Two time constants of 210 and 480 fs were fitted for the initial dynamics, and
tentatively assigned to the S$_1$/T$_2$ intersystem crossing and the
T$_2$/T$_1$ internal conversion, and near unity triplet yield was reported
\cite{mai.nc.v7.2016.p13077}. Consequently, a substantial population of the
T$_1$ ($\pi_{\rm S}\pi^{\ast}$) state was reported after the initial \mbox{400
fs}. Much less is known about deactivation of this reactive triplet state.
Generally, for all thiobases solvent quenching dominates the deactivation of
T$_1$ ($\pi\pi^{\ast}$) state in polar solvents while triplet self-quenching
is otherwise important \cite{pollum.tcc.v355.2015.p245}.

\begin{figure}[h!]
\centering\includegraphics[width=1.0\linewidth]{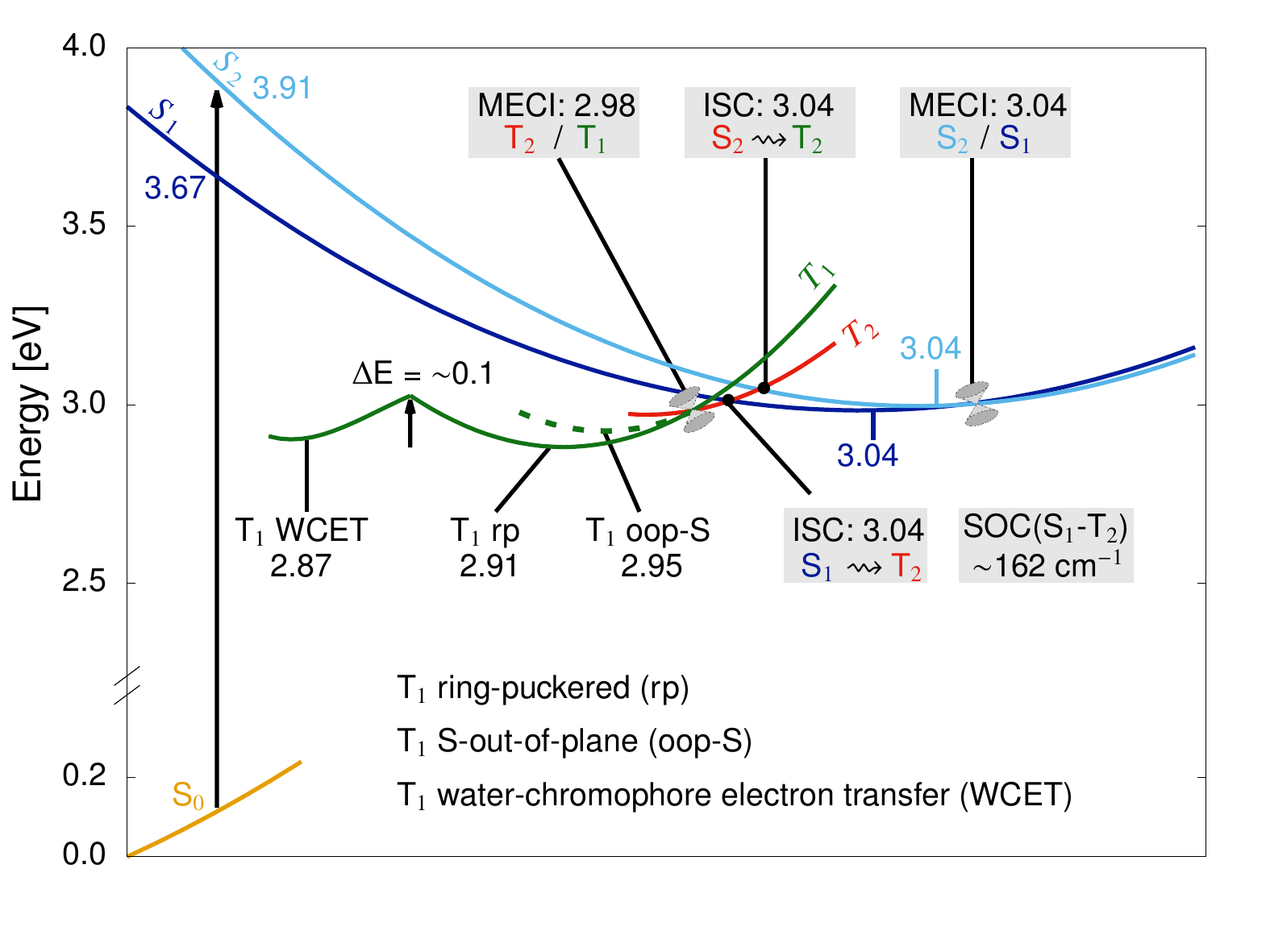}
\caption{Schematic representation of plausible photodeactivation channels in
microsolvated 2tCyt estimated at the ADC(2)/MP2/cc-pVTZ level.}
\label{fig:photochem-2tCyt}
\end{figure}

In order to further interpret the experimental TAS results
\cite{mai.nc.v7.2016.p13077} and the possible effects of explicit
solvation we optimized stationary points and identified relevant
MECPs of microhydrated 2tCyt at the ADC(2)/cc-pVTZ level. Here, we
discuss the plausible radiationless relaxation channels
(Fig.~\ref{fig:photochem-2tCyt}) upon photoexcitation at 274 nm
(4.52 eV). 
According to Kasha's rule, the initial population of optically
bright S$_4$ state (4.52 eV) should undergo ultrafast internal
conversion to the S$_2$ ($\pi\pi^{\ast}$) state. The population of
the latter state may further lead to the efficient population of the
S$_1$ state, which might be attained through the S$_2$/S$_1$ MECP at
3.04 eV. A competitive relaxation channel might drive the system
towards the T$_2$ triplet state which could be reached by
intersystem crossing in the vicinity of the
S$_2$($\pi\pi^{\ast}$)/T$_2$($n_{\rm S}\pi^{\ast}$) MECP, which is
isoenergetic with the aforementioned S$_2$/S$_1$ conical
intersection (3.04 eV). The lowest-lying T$_1$ triplet state can be
also accessed through ISC at S$_1$($\pi_{\rm
S}\pi^{\ast}$)/T$_2$($n_{\rm S}\pi^{\ast}$) MECP at 3.04 eV with SOC
of about 163 cm$^{-1}$ and the subsequent T$_2$/T$_1$ conical
intersection at 2.98 eV. In fact, the S$_1$$\rightarrow$T$_2$ ISC
should be very efficient due to the molecular orbital character
change, what is confirmed by sizable SOC matrix elements and nearly
degenerate energy levels at the stationary points lying close to the
S$_1$/T$_2$ MECP. We located ring-puckered and S-out-of-plane
minimum-energy structures having the energies of 2.91 and 2.95 eV
(with respect to the ground state geometry), on the T$_1$
hypersurface of the 2tCyt-$\ce{(H2O)2}$ cluster.
These results show that the ADC(2) picture of radiationless
deactivation of microsolvated 2tCyt system is compatible with the
previous interpretation of TAS results and MS-CASPT2 calculations
\cite{mai.nc.v7.2016.p13077}.

Interestingly, the ring-puckered T$_1$ minimum energy structure at 2.91 eV
shown in Fig.~\ref{fig:geoms}, displaying $^{3}\pi_{\rm S}\pi^{\ast}$
character, was also observed by Mai \textit{et al.}
\cite{mai.nc.v7.2016.p13077}.  However, according to our calculations, water
molecules may significantly stabilize the $^{3}n_{\rm S}\pi^{\ast}$ component
of the T$_1$ state through the water-chromophore electron transfer (WCET)
phenomenon which leads to formation of another minimum energy structure at
2.87 eV that is accessible from the ring puckered T$_1$ minimum after passing
through a modest energy barrier of \mbox{$\sim$0.1 eV}. 
This WCET process might, in fact, facilitate a radiationless deactivation
channel which ensues quenching of triplet of states in aqueous 2tCyt. Below,
we show a possible mechanism of such deactivation via a consecutive
electron-driven proton transfer mechanism.

\subsection{Water-chromophore electron transfer in microhydrated
2-thiocytosine}
Fig.~\ref{fig:geoms} shows minimum-energy structures of the ground
and lowest-lying triplet electronic states of 2tCyt-$\ce{(H2O)2}$
cluster. The $^3n_{\rm S}\pi^{\ast}$ WCET minimum is characterized
by substantial rearrangement of water molecules with respect to the
ground-state geometry.  More specifically, the water molecule which
in the ground state was hydrogen-bonded to the N3 atom of 2tCyt is
partially rotated in a way allowing attractive interaction between
its lone-pair $p_{z}$ orbital and the $n_{\rm S}$ orbital of
thiocarbonyl group (see also Tab. S2 in the SI). This
dispersion-like interaction is associated with charge transfer from
the water molecule to the chromophore which amounts to approximately
0.15 electron (according to particle-hole analysis of the electronic
wave-function at ADC(2)/cc-pVTZ level) \cite{theo}.
The corresponding distance between the sulphur and the interacting
oxygen atoms amounts to {2.40~\AA} and is noticeably shorter than
\ce{H2O}$\cdots$S distance in the electronic ground state. In
contrast, the {O--H} bond in this water molecule, interacting
through a hydrogen bond with the N3 atom of 2tCyt is extended to
{1.07~\AA}.
This structure is similar to the WCET minimum-energy crossing points
reported previously for 7H-adenine and cytosine
\cite{barbatti.jacs.v136.2014.p10246, szabla_waterchromophore_2017},
however, it is the first time that this phenomenon is found also in
the triplet manifold and with the involvement of the thiocarbonyl
group.

\begin{figure}[h!]
\centering\includegraphics[width=1.0\linewidth]{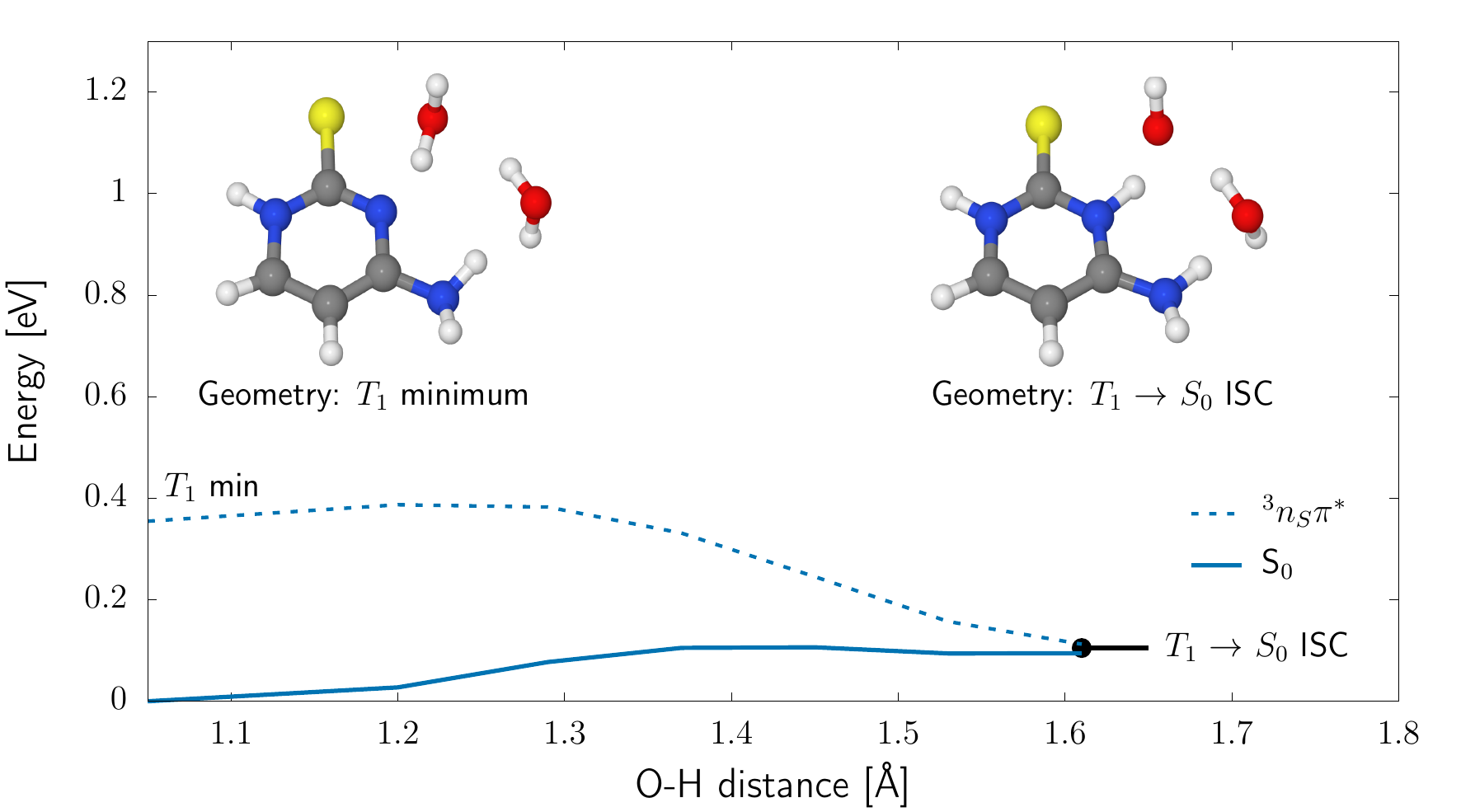}
\caption{The PE surface cuts of the lowest-lying states of microsolvated 2tCyt
showing stabilization of the $^3n_{\rm S}\pi^{\ast}$ hypersurface by WCET. The
PE profiles suggest that this process enables radiationless deactivation
channel in polar solvent that proceeds through the $^3n_{\rm
S}\pi^{\ast}$$\rightarrow$S$_0$ ISC. The plot shows linear interpolation in
internal coordinates (LIIC) between the T$_1$ WCET minimum-energy geometry and
the $^3n_{\rm S}\pi^{\ast}$$\rightarrow$S$_0$ MECP ISC, both located employing
the ADC(2)/MP2/cc-pVTZ method.}
\label{fig:2tRibCyt-cuts}
\end{figure}

The $^3n\pi^{\ast}$ minimum-energy structure lies only \mbox{0.36 eV} above
the closed-shell ground state at the ADC(2) level, and the corresponding SOC
between the $^3n\pi^{\ast}$ and S$_{0}$ amounts to \mbox{46 cm$^{-1}$}. This
indicates that 2tCyt might in fact undergo relatively efficient ISC to the
electronic ground state from this T$_{1}$ minimum. The
$^3n\pi^{\ast}$--S$_{0}$ energy gap is further reduced to 0.22 eV when more
explicit water molecules are considered (\textit{i.e.} in
[2tCyt(H$_2$O)$_4$] cluster, cf. the SI for details). Thus the observed charge
transfer and elongation of the {O--H} bond might initiate a proton transfer
from water molecule to the N3 atom of 2tCyt. According to our calculations,
such electron-driven proton transfer results in the formation of the
T$_{1}$/S$_0$ state crossing. The corresponding MECP geometry located at the
ADC(2)/MP2/cc-pVTZ level is shown as inset in Fig.~\ref{fig:2tRibCyt-cuts}.
This photochemical reaction leads to the formation of radical form of 2tCyt
and hydroxyl radical which is known to be highly reactive and mobile. This
implies that such mechanism might initiate photoinduced conversion of
2-thiocytosine to thiouracil after the subsequent migration of the hydroxyl
radical to the C4 atom of the pyrimidine ring and extrusion of ammonia. The
UV-induced deamination reaction was in fact observed experimentally in
canonical cytosine and cytidine \cite{powner.n.v459.2009.p239}.

Fig.~\ref{fig:2tRibCyt-cuts} shows potential energy profile corresponding to
the photorelaxation pathway connecting the minimum-energy structure of
$^3n_{\rm S}\pi^{\ast}$ state stabilized by WCET and the $^3n_{\rm
S}\pi^{\ast}$/S$_0$ MECP ISC, all located at the ADC(2)/MP2/cc-pVTZ level.
The shallow $^3n_{\rm S}\pi^{\ast}$ minimum visible on the left-hand side of
the PE profile actually vanishes when the water cluster is extended to several
explicit water molecules (cf. SI for details), indicating that in this case
the electron-driven proton transfer may be a virtually barrierless process.
The spin-orbit coupling (SOC) between the T$_1$ and S$_0$ states calculated
for this MECP geometry amounts to 63 $cm^{-1}$, what suggests that the
closed-shell S$_{0}$ state might be efficiently repopulated at this state
crossing. It should be noted though that the foregoing photodeactivation
channel may also lead to the formation a triplet biradical system.

To assess applicability of ADC(2) method to describe the partially biradical
$^{3}n_{\rm S}\pi^{\ast}$ state of microhydrated system, we employed the
multireference configuration interaction with single and double excitations
MR-CISD(2,3)/6-31++G(d,p) method to reoptimize its minimum-energy structure
and $^3n_{\rm S}\pi^{\ast}$/S$_0$ state crossing. Both of these MR-CISD
optimizations resulted in geometries characterized by degenerate $^3n_{\rm
S}\pi^{\ast}$ and S$_0$ states. Even though the ADC(2) results suggested a
T$_{1}$-S$_{0}$ energy gap of \mbox{0.36 eV} in the $^3n_{\rm S}\pi^{\ast}$
minimum, such $\Delta$E is in fact in the infrared spectral region and
consequently the corresponding state crossing might be accessible under
ambient conditions. Therefore, despite small quantitative differences between
the MR-CISD and ADC(2) methods, the qualitative features of the WCET minimum
are correctly reproduced at the ADC(2) level. In contrast, the $^3n_{\rm
S}\pi^{\ast}$/S$_0$ MECP associated with the proton transfer process and
optimized at the MR-CISD level is higher in energy by 0.43 eV than the
corresponding T$_1$ WCET minimum. This suggests lower availability of the EDPT
process than that implied by the ADC(2) method. 

\subsection{Simulations of excited-state spectra of 2-thiocytosine}
\begin{figure}[htpb!]
\centering\includegraphics[width=0.8\linewidth]{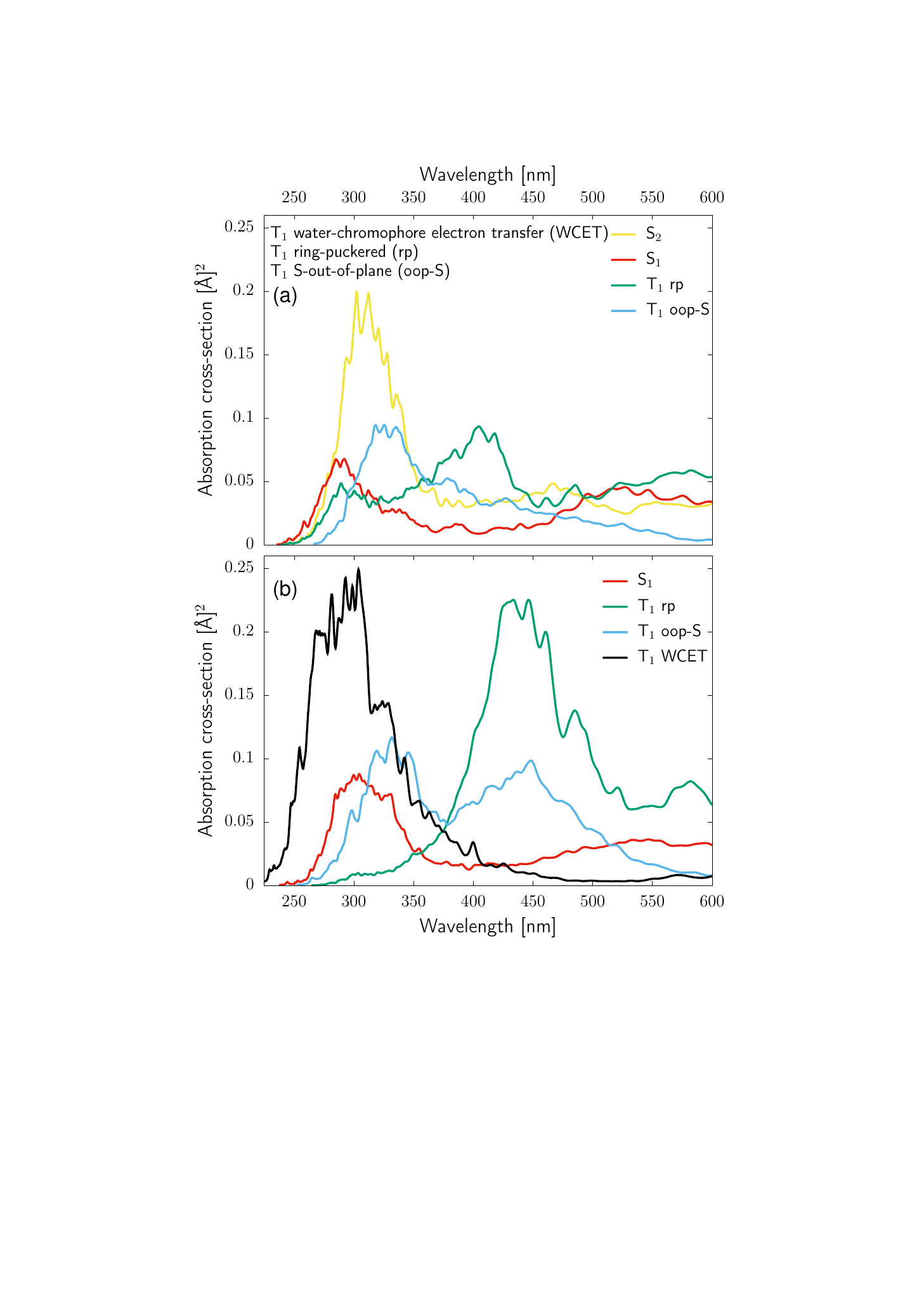}
\caption{Excited-state absorption (ESA) spectra for the isolated 2tCyt (a) and
microsolvated 2tCyt (b) at the ADC(2)/cc-pVTZ level.}
\label{fig:ESA}
\end{figure}

In order to verify our findings and compare our results to previous
experimental and theoretical results of Mai at al.
\cite{mai.nc.v7.2016.p13077}, we performed simulations of excited-state
absorption (ESA) spectra for isolated 2tCyt (Fig. \ref{fig:ESA}a) and
microhydrated 2tCyt (Fig. \ref{fig:ESA}b) molecules employing the nuclear
ensemble method.
The simulated ESA spectra can be further used in assignment of transient
absorption (TAS) UV measurements. The simulated ESA spectra for the isolated
molecule were obtained using the minimum-energy structures corresponding to the
S$_1$, S$_2$, T$_1$ ring-puckered and T$_1$ S-out-of-plane minima optimized at
the ADC(2)/cc-pVTZ level, whereas in the case of microhydrated molecules the
ESA spectra were simulated assuming S$_1$, T$_1$ ring-puckered, T$_1$
S-out-of-plane and T$_1$ WCET excited state potential energy surfaces.

Comparison of the ESA spectra (Fig. \ref{fig:ESA}) obtained for both isolated
and microhydrated 2tCyt shows similar features with somewhat redishifted
absorption bands for the latter.
According to the experimental TAS spectrum of 2tCyt in water solution
\cite{mai.nc.v7.2016.p13077}, there are two distinct absorption maxima at
about 350 and 525 nm which were assigned based on MS-CASPT2 calculations to
the $^1\pi_{\rm S}\pi^{\ast}$ and $^3\pi_{\rm S}\pi^{\ast}$ states,
respectively.  Our simulated ESA spectra (Fig. \ref{fig:ESA}a) of 2tCyt are
generally consistent with these findings and show that these bands may refer
to the S$_2$ (or S$_1$) and T$_1$  states in ADC(2) calculations,
respectively, for which the computed absorption maxima are located at about
310 and 400 nm, respectively. It is worth noting that the latter state (T$_1$
ring-puckered minimum) has two components, mixing the $^3\pi_{\rm S}\pi^{\ast}$
and $^3n_{\rm S}\pi^{\ast}$ transitions, in line with what was suggested in the
previous work conducted at the MS-CASPT2 level \cite{mai.nc.v7.2016.p13077}.
The disparity between experimental and simulated maxima (310 nm vs 350 nm)
assigned to the singlet $^1\pi_{\rm S}\pi^{\ast}$ state (S$_2$ or S$_1$) may
be due to lack or inappropriate description of solvent effects in the
simulated ESA spectrum. It should be noted though that our ESA spectra are
simulated from the corresponding minimum energy geometries distorted along
vibrational normal modes, and thus do not correspond to fully relaxed PE
surfaces.
Even more pronounced differences in the experimental and theoretical
absorption bands (400 nm compared to 525 nm) are apparent for the T$_{1}$
ring-puckered state. However, vibrational cooling effects result in systematic
blue-shift of this band during the excited state dynamics of the system and
the position of the experimental maximum (525 nm) was reported at \mbox{3.7
ps}. Consequently, we expect this band to undergo further hypsochromic shift
at longer time delays. 

The ESA spectrum simulated for the T$_1$ ring-puckered state ($^3\pi_{\rm
S}\pi^{\ast}$) of microhydrated  2tCyt-\ce{(H2O)2} reveals slight redshift of
the predicted absorption maximum to 450 nm. In addition, the simulated ESA
spectrum (Fig. \ref{fig:ESA}b) of the T$_1$ S-out-of-plane ($^3\pi_{\rm
S}\pi^{\ast}$/$^3n_{\rm S}\pi^{\ast}$) minimum-energy structure exhibits an
overlapping band at 450 nm, similar to the ESA spectrum of the T$_1$
ring-puckered state. This indicates that both the T$_1$ ring-puckered and
T$_1$ S-out-of-plane minima could be populated during the excited-state
dynamics of aquated 2tCyt.
As shown in Fig.~\ref{fig:ESA}b the simulated ESA spectrum for the
T$_1$($^3n_{\rm S}\pi^{\ast}$) WCET minimum is characterized by high
absorption in the same spectral range as the S$_{1}$ and S$_{2}$ states of
both microhydrated and isolated 2tCyt. Consequently, the absorption band
visible at 350 nm between 2 and 4 ps in the experimental TAS spectrum might be
also the result of partial population of the T$_1$($^3n_{\rm S}\pi^{\ast}$)
WCET state \cite{mai.nc.v7.2016.p13077}. Unfortunately, the corresponding
excited-state absorption band overlaps with that of the T$_1$ S-out-of-plane
with dominant contribution from the $^3\pi_{\rm S}\pi^{\ast}$ configuration.
Therefore, analysis of TAS measurements is probably not sufficient to provide
unambiguous identification of the different states contributing to the
excited-state absorption in the triplet manifold. 

It is worth noting, that the WCET process is characterized by very
small T$_{1}$/S$_{0}$ energy gaps and relatively high SOC matrix
elements (46 cm$^{-1}$) what suggests that it might not be
observable in TAS measurements due to rather short lifetimes.
However, population of the $^3\pi_{\rm S}\pi^{\ast}$ WCET state
might lead to efficient proton transfer and formation of the
hydroxyl radical or even unreactive repopulation of the electronic
ground state. Another scenario induced by the characteristic WCET
interaction may be related to the population of the S-out-of-plane
T$_{1}$ minimum, which according to our preliminary XMS-CASPT2
optimizations has sufficient contribution from the $^3n_{\rm
S}\pi^{\ast}$ state to induce the characteristic \ce{H2O}$\cdots$S=C
interaction (cf. SI for details). In each of these cases, we expect
that the WCET photodeactivation channel described in detail in the
previous paragraphs could efficiently quench triplet states of 2tCyt
with the direct involvement of solvent molecules. Consequently the
WCET process could explain the weak photosensitizing properties of
aqueous 2tCyt solutions.

\section{Conclusions}

In order to elucidate radiationless deactivation of 2tCyt in aqueous
environment, we performed computational explorations of
excited-state potential energy surfaces of microhydrated 2tCyt model
system. We discovered that water-chromophore electron transfer
(WCET) might be the potential driving force responsible for
quenching excited triplet states of the studied molecule.
For the first time, we show that this characteristic WCET
interaction may involve a thiocarbonyl group leading to an
interaction between the $p_{z}$ orbital of water and the $n_{\rm S}$
orbital of the chromophore. Furthermore, this is the first example
of a WCET process occurring in the triplet manifold of electronic
states \cite{szabla_waterchromophore_2017,
barbatti.jacs.v136.2014.p10246}. In particular, our results
demonstrate that the T$_{1}$ topography of aqueous thionucleobases
cannot be limited to the sole consideration of S-out-of-plane and
ring-puckered $^{3}\pi\pi^{\ast}$ minima (as suggested by Bai
\textit{et al.}) \cite{bai_decay_2017}, and in specific cases
$^{3}n\pi^{\ast}$ electronic states might come into play. As shown
above, very low T$_{1}$-S$_{0}$ energy gap and considerable SOC
matrix elements could provide an efficient channel for the
repopulation of the closed-shell electronic ground state.

Since photoinduced electron transfer processes often entail subsequent proton
transfer, we considered similar possibility from the WCET minimum of
microhydrated 2tCyt. According to our ADC(2) simulations this virtually
barrierless EDPT process might be resulting in the formation of hydroxyl
radical which could further participate in photohydration or deamination
reactions.  It should be noted though that the additional MR-CISD results
suggest existence of a modest energy barrier (below 0.5 eV) for this process.
In fact, this phenomenon is closely related to UV-induced water splitting
reactions observed for pyridine \cite{liu_computational_2013,
liu_photoinduced_2014}, acridine \cite{liu_photocatalytic_2015} and heptazine
\cite{ehrmaier_mechanism_2017}. Therefore, we suggest that further studies of
thionated compounds with similar photochemical properties might be an
interesting further direction for photochemical water splitting, particularly
since, the aforementioned nitrogenous heterocycles exhibit predominantly
singlet and not triplet reactive photochemistry.

Resuming, our study demonstrates that water molecules may modify the
shapes of excited-state PE surfaces in thionucleobases, stabilize
exciplex interactions and open photorelaxation channels which are
not available in isolated chromophores.  Since these effects cannot
be reproduced theoretically by sole use of implicit solvation models
or hybrid quantum mechanics/molecular mechanics (QM/MM) approach, we
suggest that the inclusion of explicit QM water molecules might be
the necessary next step in determining the important details of the
photochemistry of aqueous thionucleobases.

\section*{Supplementary data}
Supplementary data associated with this article can be found, in the
online version including: vertical excitation energies of
microhydrated 2-thiocytosine (2tCyt) in gas phase; detailed
discussion of the T$_1$ minimum energy structures of microhydrated
2-thiocytosine (2tCyt) located using various methods; properties of
larger microhydrated clusters and the results of QM/EFP
calculations.

\section*{Acknowledgments}
This work was supported by a fellowship from the Simons Foundation
(494188 to R.S.), financial support from the National Science Centre
Poland (2016/23/B/ST4/01048 to R.W.G.) and a statutory activity
subsidy from the Polish Ministry of Science and Higher Education for
the Faculty of Chemistry of Wroclaw University of Science and
Technology. M.J.J. acknowledges support within the ``Diamond Grant''
(0144/DIA/2017/46) from the Polish Ministry of Science and Higher
Education. Computational grants from Interdisciplinary Centre for
Mathematical and Computational Modelling (ICM, Grant No. G53-28) and
Wrocław Centre of Networking and Supercomputing (WCSS) are also
gratefully acknowledged.

\section*{References}

\bibliography{refs.thioC.2017}

\end{document}